\documentclass[12pt]{iopart}

\usepackage{iopams} 
\usepackage{bm,graphics,graphicx,epsfig,color}

 \newcommand{\apj}{Astrophys.\ J.}
 
  \newcommand{\mnras}{Mon.\ Not.\ R.\ Astron.\ Soc.}
  \newcommand{\prd}{Phys.\ Rev.\ D}

     \newcommand{\prl}{Phys.\ Rev.\ Lett.}

 \newcommand{\cqg}{Class.\ Quantum Grav.}
 \newcommand{\lrr}{Living Rev.\ Relativ.}
 
  \newcommand{\pla}{Phys.\ Lett.\ A}
  \newcommand{\apss}{Astrophys.\ Sp.\ Sci.}
  \newcommand{\ijtp}{Int.\ J.\ Theor.\ Phys.}

\def\v1v2{{\bf v}_1 \cdot {\bf v}_2}
\def\tp{\tilde{p}}
\def\te{\tilde{e}}
\def\tal{\tilde{\alpha}}
\def\tbe{\tilde{\beta}}
\def\tom{\tilde{\omega}}

\begin{document}

\title[Pericenter advance]{Pericenter advance in general relativity: Comparison of approaches at high post-Newtonian orders}

\author{Alexandria Tucker$^1$ and Clifford M. Will$^{1,2}$}

\ead{a.tucker@ufl.edu, cmw@phys.ufl.edu}
\address{
$^1$ Department of Physics,
University of Florida, Gainesville FL 32611, USA
 \\
$^2$ GReCO, Institut d'Astrophysique de Paris, UMR 7095-CNRS,
Sorbonne Universit\'e, 98$^{bis}$ Bd. Arago, 75014 Paris, France
}

\date{\today}

\begin{abstract}
The advance of the pericenter of the orbit of a test body around a massive body in general relativity can be calculated in a number of ways.   In one method, one studies the geodesic equation in the exact Schwarzschild geometry and finds the angle between pericenters as an integral of a certain radial function between turning points of the orbit.   In another method, one describes the orbit using osculating orbit elements, and analyzes the ``Lagrange planetary equations'' that give the evolution of the elements under the perturbing effects of post-Newtonian (PN) corrections to the motion.  After separating the perturbations into periodic and secular effects, one obtains an equation for the secular rate of change of the pericenter angle.  While the different methods agree on the leading post-Newtonian contribution to the advance, they do not agree on the higher-order PN corrections.  We show that this disagreement is illusory.  When the orbital variables in each case are expressed in terms of the invariant energy and angular momentum of the orbit and when account is taken of a subtle difference in the meaning of ``pericenter advance'' between the two methods, we show to the third post-Newtonian order that the different methods actually agree perfectly.  

\end{abstract}

\maketitle

\section{Introduction and summary}
\label{sec:intro}  

The advance of the pericenter of the orbit of a two-body system is an iconic general relativistic effect.  The observed anomaly in the perihelion advance of Mercury was used by Einstein to test his various drafts of a theory of gravity, and became an empirical cornerstone of the final theory that he presented in November 1915.  In binary pulsar systems, measurements of the pericenter advance combined with assuming the validity of general relativity have been used to measure the total mass of the systems, and have helped to provide strong-field tests of the theory (for reviews see \cite{tegp2,2014LRR....17....4W}).

These observations rely only upon the leading contribution predicted by general relativity in a post-Newtonian (PN) expansion, which is an approximation to the theory which treats $GM/rc^2 \sim (v/c)^2$ as small quantities, where $M$, $r$, and $v$ are characteristic masses, separations and velocities in the system, and $G$ and $c$ are the gravitational constant and speed of light, respectively (for a pedagogical review of post-Newtonian theory, see \cite{PW2014}).  For a two-body system, that leading contribution is 
\begin{equation}
\Delta \omega = \frac{6\pi GM}{c^2 p} \,,
\label{eq:delomega0}
\end{equation}
where $\Delta \omega$ is the change in the pericenter angle over one orbit, $M$ is the total mass of the system, and $p = a(1-e^2)$ is the semilatus rectum, where $a$ and $e$ are the semimajor axis and eccentricity of the orbit.

Despite numerous studies of higher-order PN corrections to the pericenter advance \cite{1976PThPh..56..324H,1988NCimB.101..127D,1989PThPh..81..679O,1998PhLA..238..328D,2009IJTP...48.2198H,2011Ap&SS.332..121D,2014LRR....17....2B,2018MNRAS.480.3747W,2018Ap&SS.363..245M},
such corrections have not been important observationally to date.  For Mercury, they are smaller than the leading term by a factor of $10^8$, too small to be detectable, even by the recently launched BepiColombo mission to send two orbiters around Mercury \cite{2018PhRvL.120s1101W}.  Future observations of the ``double pulsar'' system J0737-3039 could begin to be sensitive to the 2PN contribution to $\Delta \omega$ \cite{2009CQGra..26g3001K}.   Such corrections could also be relevant when ground based gravitational-wave observatories begin to detect compact binary inspirals with significant residual eccentricity, and will undoubtedly be relevant for detection of waves from high-eccentricity extreme mass-ratio inspirals by space-based detectors such as LISA.  There are also suggestions that incorporating such higher-order PN corrections may be important for testing the massive black hole binary model for the blazar OJ 287 \cite{2018ApJ...866...11D}.

It is routine for general relativity textbooks to derive the pericenter advance (\ref{eq:delomega0}), at least for the special case of a test particle undergoing geodesic motion in the Schwarzschild metric \cite{Weinberg,1973grav.book.....M,1984ucp..book.....W,2003gieg.book.....H,2004sgig.book.....C,2004graa.book.....S,2009fcgr.book.....S,PW2014}.  Making use of the existence of a conserved total mass-energy $E$ and angular momentum $L$, one can reduce the geodesic equation to an ordinary differential equation for the radius $r$ in terms of the angle $\phi$.  The angle between successive turning points, or extrema of $r$, can be obtained exactly from a radial integral.   The expansion of that result in a post-Newtonian sequence yields Eq.\ (\ref{eq:delomega0}) at lowest order. 

However, for more general orbital problems, such as binary orbits where the masses are comparable, where tidal effects or spin effects must be included, or where there are perturbations due to additional bodies, more powerful and general techniques are available for calculating such relativistic effects.  One of these is the method of ``osculating orbit elements'', whereby one characterizes a binary orbit by an instantaneous Keplerian orbit, whose orbital elements evolve with time as a result of the perturbing forces.

In this paper we address the curious fact that, while the ``textbook'' method and the osculating orbit method agree on the pericenter advance at lowest PN order, they seem to give different results for the higher-order PN corrections, even when they are carried out in the same coordinate system.   However, we will show that,  
when the orbital variables in each case are related to the invariant energy and angular momentum of the orbit and when account is taken of a subtle difference in the meaning of Òpericenter advanceÓ between the two methods, they are in complete agreement.
 
For example, carrying out the textbook method using the Schwarzschild geometry in harmonic coordinates and expanding the result to third post-Newtonian order gives the pericenter advance
\begin{equation}
\Delta \omega =  6\pi \frac{m}{ p} + \frac{3\pi}{2} \left (\frac{m}{p} \right )^2 (14- 3e^2) + \frac{3\pi}{2}  \left (\frac{m}{ p} \right )^3 (58-13e^2) \,,
\label{eq:deltaomh}
\end{equation}
where $m \equiv GM/c^2$ and $p$ and $e$ are defined by the pericenter and apocenter radii by
\begin{equation}
r_- \equiv \frac{p}{1+e} \,, \quad r_+ \equiv \frac{p}{1-e} \,.
\label{eq:turning0}
\end{equation} 

In the osculating elements method, one
begins with the 3PN equations of motion for a test body orbiting a spherical body of mass $M$, also in harmonic coordinates.   In this approach, the orbital separation is defined by 
\begin{equation}
r \equiv \frac{p}{1+e\cos(\phi - \omega)} \,,
\end{equation}
where $p$, $e$ and $\omega$ are osculating orbit elements that may vary with time.
One then uses a two-timescale analysis that separates the evolution of the osculating orbit elements into terms that vary on an orbital timescale and those that vary on a long, secular timescale, and finds the ``orbit-averaged'' rate of advance of the osculating pericenter angle,
\begin{equation}
\frac{d\tom}{d\phi} = 3 \frac{m}{\tp} - \frac{3}{4} \left ( \frac{m}{ \tp} \right )^{2} \left ( 10 - \te^2  \right )
+ \frac{3}{2} \left ( \frac{m}{ \tp} \right )^{3}  \left (29 + 34 \te^2 \right )  \,,
\label{eq:domdtheta2} 
\end{equation}
where $\tp$ and $\te$ are the orbit-averaged osculating semilatus rectum and eccentricity, respectively.  They are constant over the long timescale.  

While Eq.\ (\ref{eq:deltaomh}) and $2\pi$ times Eq.\ (\ref{eq:domdtheta2}) agree to 1PN order, they clearly disagree at higher orders.
But this disagreement is illusory, because the semilatus rectum and eccentricity have different meanings in the two approaches.   However, in each approach the motion is characterized by a conserved total energy $E$ and angular momentum $L$ (both per unit mass of the test particle), which are invariant quantities (each is the inner product between the four-velocity and a Killing vector associated with the symmetry of the metric), and are expressible in terms of the semilatus rectum and eccentricity corresponding to each method.   These relations can be inverted to express both orbital variables  in terms of $E$ and $L$.  

One way to do this (though not the only way) is to 
define the dimensionless angular momentum 
$L^\dagger \equiv cL/GM$,
and then define the ``invariant'' semilatus rectum $p_0$ and eccentricity $e_0$ according to
\begin{eqnarray}
p_0 &\equiv& \frac{GM}{c^2} (L^\dagger)^2 \,,
\nonumber \\
e_0^2 &\equiv& 1 + (L^\dagger)^2 \left ( \frac{E^2}{c^4} -1 \right ) \,.
\label{eq:epdefinitions}
\end{eqnarray}
In the Newtonian limit, these correspond to the conventional definitions.  Then each pair $(p,\,e)$  and $(\tp ,\, \te)$ can be related to $(p_0 ,\, e_0)$ in a PN expansion.   

However, we must also take into account the fact that, in the osculating orbit method, $2\pi (d\tom/d\phi)$ {\em is not equal to the angle between successive pericenters}.  This can be seen as follows:  including only the secular effects on the osculating orbit elements, the radial separation is defined by
\begin{equation}
r = \frac{\tp}{1 + \te \cos (\phi - \omega_0 - \phi \, d\tom/d\phi )} \,,
\end{equation}
where $\te$, $\tp$ and $d\tom/d\phi$ are constant and $\omega_0$ is the initial pericenter angle.  Pericenter occurs when the argument of the cosine is $2\pi N$, where $N$ is an integer.  Successive pericenters then occur when $\Delta \phi (1 -  d\tom/d\phi) = 2\pi$.  Therefore the advance of the pericenter angle over each orbit is given by $\Delta \omega = \Delta \phi - 2\pi$, or
\begin{equation}
\Delta \omega = \frac{2 \pi d\tom/d\phi}{1-d\tom/d\phi} \,.
\label{eq:delomega}
\end{equation}
At 1PN order the denominator does not contribute, but at higher orders, it does, and with this effect incorporated, the pericenter advance in the osculating method now agrees with that from the textbook method.  
Expressed in terms of the ``invariant'' $p_0$ and $e_0$ both methods give the same pericenter advance
\begin{equation}
\Delta \omega = 6\pi \frac{m}{ p_0} + \frac{15\pi}{2} \left (\frac{m}{p_0} \right )^2 (6+e_0^2) + \frac{105\pi}{2}  \left (\frac{m}{p_0} \right )^3 (8+3e_0^2) \,.
\label{eq:deltaominv}
\end{equation}
This illustrates the importance of expressing results in general relativity in terms of invariant or observable quantities.

The remainder of this paper gives the details leading to these conclusions.  In Sec.\ \ref{sec:geodesic}, we analyse the pericenter advance using the exact Schwarzschild geometry, working in both Schwarzschild and harmonic coordinates.  In Sec.\ \ref{sec:eom} we use the method of osculating orbit elements.  Section \ref{sec:concluding} makes concluding remarks.

\section{The textbook method}
\label{sec:geodesic}

\subsection{Schwarzschild coordinates}

We begin with the Schwarzschild metric in spherical Schwarzschild coordinates, 
\begin{equation}
ds^2 = -\left ( 1-\frac{2m}{ r_{\rm s}} \right ) dt^2 + \left ( 1-\frac{2m}{ r_{\rm s}} \right )^{-1} dr_{\rm s}^2 + r_{\rm s}^2 (d\theta^2 + \sin^2 \theta d\phi^2 ) \,.
\label{eq:metric1}
\end{equation}
Because the metric is static and spherically symmetric, we can place the orbit of a test particle on the equatorial plane ($\theta = \pi/2$), and write down the conserved energy and angular momentum per unit mass of the test particle, 
\begin{eqnarray}
E &=& c^2(1 - 2m/ r_{\rm s} ) dt/d\tau \,,
\nonumber \\
L &=& r_{\rm s}^2 d\phi/d\tau \,,
\label{eq:EL1}
\end{eqnarray}
where $\tau$ is proper time along the world line of the body.  From the normalization of the four-velocity, $u^\alpha u^\beta g_{\alpha\beta} =-c^2$, we obtain the radial equation
\begin{equation}
 \left ( \frac{dr_{\rm s}}{d\tau} \right )^2 = \frac{E^2}{c^2} - c^2 \left (1-\frac{2m}{r_{\rm s}} \right ) \left (1 + \frac{L^2}{c^2 r_{\rm s}^2}  \right ) \,.
\label{eq:drdt2}
\end{equation}
We define the inner and outer turning points of the orbit where $dr_{\rm s}/d\tau = 0$ (pericenter and apocenter, respectively) by
\begin{equation}
(r_{\rm s})_- \equiv \frac{p_{\rm s}}{1+e_{\rm s}} \,, \quad (r_{\rm s})_+ \equiv \frac{p_{\rm s}}{1-e_{\rm s}} \,,
\label{eq:turningS}
\end{equation}
and recast Eq.\ (\ref{eq:drdt2}) in the form
\begin{equation}
\frac{1}{2} \left ( \frac{dr_{\rm s}}{d\tau} \right )^2 = mL^2 \left (\frac{1}{r_{\rm s}} - \frac{1}{r_0} \right )\left (\frac{1}{r_{\rm s}} - \frac{1}{(r_{\rm s})_-} \right )\left (\frac{1}{r_{\rm s}} - \frac{1}{(r_{\rm s})_+} \right ) \,,
\label{eq:drdt3}
\end{equation}
where $r_0 = 2m/(1-4m/p_{\rm s})$ (see eg.\ Sec.\ 5.6.3 of \cite{PW2014}).   The energy and angular momentum are related to $m$, $p$, and $e$ by
\begin{eqnarray}
\frac{E^2}{c^2} &=& c^2 - \frac{GM(1-e_{\rm s}^2)}{p_{\rm s}} \frac{1-4m/p_{\rm s}}{1-(3+e_{\rm s}^2)m/p_{\rm s}} \,,
\nonumber \\
L^2 &=& \frac{GMp_{\rm s}}{1-(3+e_{\rm s}^2)m/p_{\rm s}} \,.
\label{eq:EL2}
\end{eqnarray}
Converting Eq.\ (\ref{eq:drdt3}) from $d/d\tau$ to $d/d\phi$ using the second of Eqs.\ (\ref{eq:EL1}), and making the change of variables
\begin{equation}
r_{\rm s}(\chi) = \frac{p_{\rm s}}{1 + e_{\rm s} \cos \chi } \,,
\end{equation}
where $\chi$ runs from $0$ (where $r_{\rm s}=(r_{\rm s})_-$) to $\pi$ (where $r_{\rm s}=(r_{\rm s})_+$) and then to $2\pi$, we obtain
\begin{equation}
\left ( \frac{d\chi}{d\phi} \right )^2 = 1 - \frac{2m}{p_{\rm s}} \left (3+ e_{\rm s} \cos \chi \right ) \,.
\end{equation}
Then as $\chi$ ranges from $0$ to $2\pi$, i.e. from one pericenter to the next pericenter, $\phi$ ranges over an angle given by
\begin{equation}
\Delta \phi = 2 \int_0^\pi \left (1 -  \frac{2m}{p_{\rm s}} \left (3+ e_{\rm s} \cos \chi \right )\right )^{-1/2} d\chi\,.
\end{equation}
Factoring out the $\chi$-independent term $1 - 6m/p_{\rm s}$, expanding the square root in a power series and integrating over 
$\chi$, we obtain the infinite series (\cite{2018MNRAS.480.3747W}, for an alternative series, see \cite{2009IJTP...48.2198H})
\begin{equation}
\Delta \phi = \frac{2\pi}{\sqrt{1-6m/p_{\rm s}}} \left [ 1 + \sum_{n=1}^\infty \frac{(4n)!}{n!^2 (2n)! 2^{6n}} \beta^{2n} \right ] \,,
\label{eq:deltaphi}
\end{equation}
where
\begin{equation}
\beta \equiv \frac{2e_{\rm s}m/p_{\rm s}}{1-6m/p_{\rm s}} \,.
\end{equation}
The pericenter advance over one orbit is then given by $\Delta \omega = \Delta \phi - 2\pi$.  Expanding Eq.\ (\ref{eq:deltaphi}) to 3PN order, we obtain
\begin{equation}
\Delta \omega = 6\pi  \frac{m}{p_{\rm s}} + \frac{3\pi}{2} \left (\frac{m}{p_{\rm s}} \right )^2 (18+e_{\rm s}^2) + \frac{45\pi}{2}  \left (\frac{m}{p_{\rm s}} \right )^3 (6+e_{\rm s}^2) \,.
\label{eq:deltaom}
\end{equation}
Equations (\ref{eq:EL2}) can now be iterated to express $e_{\rm s}$ and $p_{\rm s}$ in terms of $E$ and $L$, and thence in terms of $e_0$ and $p_0$ via Eqs.\ (\ref{eq:epdefinitions}), with the result
\begin{eqnarray}
e_{\rm s}^2 &=& e_0^2 + 2 x_0 (1- e_0^4) + x_0^2 (1- e_0^2)(9+ 10e_0^2 - 3 e_0^4 ) + O(x_0^3) \,,
\nonumber \\
x_{\rm s} &=& x_0 + x_0^2 (3+ e_0^2) + 4x_0^3 (5 + 12e_0^2) + O(x_0^4) \,,
\label{eq:extransform1}
\end{eqnarray}
where $x_{\rm s} \equiv m/p_{\rm s}$ and $x_0 \equiv m/p_0$.  Substituting Eqs.\ (\ref{eq:extransform1}) into (\ref{eq:deltaom}) and expanding to 3PN order yields Eq.\ (\ref{eq:deltaominv}).

\subsection{Harmonic coordinates}

The harmonic radial coordinate is related to Schwarzschild coordinates by $r = r_{\rm s} - m$, and puts the  Schwarzschild metric into the form
\begin{eqnarray}
ds^2 &=& -\left ( \frac{1-m/ r}{1+m/ r} \right ) dt^2 
+ \left (  \frac{1+m/ r}{1-m/ r} \right ) dr^2 
\nonumber \\
&& \qquad
+ \left( r + m \right)^2 (d\theta^2 + \sin^2 \theta d\phi^2 ) \,.
\label{eq:metric2}
\end{eqnarray}   
Now, $E$ and $L$ are given by
\begin{eqnarray}
E &=& c^2 \left (\frac{1 - m/r}{1 + m/r} \right ) dt/d\tau \,,
\nonumber \\
L &=&  \left( r + m \right)^2 d\phi/d\tau \,,
\label{eq:EL1h}
\end{eqnarray}
and the radial equation is
\begin{equation}
 \left ( \frac{dr}{d\tau} \right )^2 = \frac{E^2}{c^2}  -c^2 \left ( \frac{1-m/r}{1+m/r} \right ) \left ( 1 +\frac{L^2}{c^2 (r + m)^2} \right ) \,.
\label{eq:drdt2h}
\end{equation}
We now define the turning points
\begin{equation}
(r)_- \equiv \frac{p}{1+e} \,, \quad (r)_+ \equiv \frac{p}{1-e} \,,
\label{eq:turningH}
\end{equation}
and repeat the procedure above, obtaining
\begin{eqnarray}
\frac{E^2}{c^2} &=& c^2 - \frac{GM(1-e^2)}{p} \left [\frac{1-2x - 3x^2 (1-e^2)}{1-2x e^2 - 3x^2 (1-e^2)- 2x^3 (1-e^2)^2} 
\right ] \,,
\nonumber \\
L^2 &=& GMp \left [\frac{\left [1+2x + x^2 (1-e^2) \right ]^2}{1-2x e^2 - 3x^2 (1-e^2)- 2x^3 (1-e^2)^2} 
\right ] \,,
\label{eq:EL2h}
\end{eqnarray}
where $x \equiv m/p$.  Making the change of variables
\begin{equation}
r(\chi) = \frac{p}{1 + e \cos \chi } \,,
\end{equation}
we finally obtain the change in the angle $\phi$ from one pericenter to the next, given by
\begin{equation}
\Delta \phi = 2 \kappa^{1/2} \int_0^\pi \left (1 -  e x \beta \cos \chi \right )^{-1/2} 
\left (1 +  \frac{e x}{1+x} \cos \chi \right )^{-1/2} d\chi\,,
\label{eq:deltaphih}
\end{equation}
where
\begin{eqnarray}
\kappa &\equiv& \frac{\left (1+2x + x^2 (1-e^2) \right )^2 }{(1+x) \left ( 
1 - 3x - 3x^2 (3 - e^2 ) -5x^3 (1-e^2) \right )} \,,
\nonumber \\
\beta &\equiv&  \frac{1+6x + 5x^2 (1-e^2)  }{\left ( 
1 - 3x - 3x^2 (3 - e^2 ) -5x^3 (1-e^2) \right )} \,.
\end{eqnarray}
Expanding Eq.\ (\ref{eq:deltaphih}) to 3PN order and carrying out the integrations, we obtain Eq.\ (\ref{eq:deltaomh}).

Iterating Eqs.\ (\ref{eq:EL2h}) we express $e$ and $p$ in terms of  $e_0$ and $p_0$ via Eqs.\ (\ref{eq:epdefinitions}), with the result
\begin{eqnarray}
e^2 &=& e_0^2 + 2 x_0 (1- e_0^2)(1+ 2e_0^2) + x_0^2 (1- e_0^2)(13+ 15e_0^2 - 12 e_0^4 ) + O(x_0^3) \,,
\nonumber \\
x &=& x_0 + 2x_0^2 (2+ e_0^2) + x_0^3 (29 + 23e_0^2) + O(x_0^4) \,.
\label{eq:extransformh}
\end{eqnarray}
Substituting Eqs.\ (\ref{eq:extransformh}) into (\ref{eq:deltaomh}) and expanding to 3PN order again yields Eq.\ (\ref{eq:deltaominv}).

\section{The post-Newtonian osculating orbits method} 
\label{sec:eom}

The equation of motion of a test body in the field of a body of mass $M$ is given, to 3PN order, by \cite{2014LRR....17....2B}
\begin{eqnarray}
\frac{d{\bm v}}{dt} &=& - \frac{GM{\bm n}}{r^2} 
-  \frac{GM}{c^2 r^2} \left [  \left ( v^2 - 4\frac{GM}{r} \right ){\bm n} - 4 \dot{r} {\bm v} \right ]  
\nonumber \\
&&
\quad -  \frac{G^2 M^2}{c^4 r^3}   \left [ \left( \frac{9GM}{r} - 2\dot{r}^2 \right ) {\bm n} + 2 \dot{r} {\bm v} \right ]
\nonumber \\
&&
\quad +  \frac{G^3 M^3}{c^6 r^4}  
\left [ \left( \frac{16GM}{r} - \dot{r}^2 \right ) {\bm n}
+ 4 \dot{r} {\bm v} \right ]
\,.
\label{eq:eom2}
\end{eqnarray}
This equation is expressed in harmonic coordinates, which are the basis for modern post-Newtonian theory \cite{PW2014}.
We apply standard orbital perturbation theory, used to compute deviations from Keplerian two-body motion induced by perturbing forces, described by the equation of motion
$d^2 {\bm x}/dt^2 = - GM{\bm n}/r^2 + \delta {\bm a}$,
where $\delta {\bm a}$ is a perturbing acceleration.  
For a general orbit described by $\bm x$ and $\bm v = d{\bm x}/dt$, we define the ``osculating'' Keplerian orbit using a set of orbit elements, the semilatus rectum $p$, eccentricity $e$ and pericenter angle $\omega$.  They are {\em defined} by the following set of equations:
\begin{eqnarray}
{\bm x} &\equiv& r {\bm n} \,,
\nonumber \\
r &\equiv& p/(1+\alpha \cos \phi + \beta \sin \phi) \,,
\nonumber \\
{\bm n} &\equiv&  {\bm e}_X \cos \phi   
 +  {\bm e}_Y \sin \phi  \,,
\nonumber \\
{\bm \lambda} &\equiv& \partial {\bm n}/\partial \phi \,, \quad \hat{\bm h}={\bm n} \times {\bm \lambda} \,,
\nonumber \\
{\bm h} &\equiv& {\bm x} \times {\bm v} \equiv \sqrt{GMp} \, \bm{\hat{h}} \,,
\label{eq2:keplerorbit}
\end{eqnarray}
where  ${\bm e}_A$ are chosen reference basis vectors, $\phi$ is the orbital phase measured from the $X$ axis, and $\alpha$ and $\beta$ are given by
 \begin{eqnarray}
\alpha &=& e \cos \omega \,,
\nonumber \\
\beta &=& e \sin \omega \,. 
\label{eq2:alphabeta}
\end{eqnarray} 
From the given definitions, we see that ${\bm v} = \dot{r} {\bm n} + (h/r) {\bm \lambda}$.  Exploiting the spherical symmetry of the problem we have chosen the orbital plane to lie on the $X-Y$ plane, so that the other orbital elements, inclination and angle of nodes, are not relevant.   
We define the radial $\cal R$, cross-track $\cal S$ and out-of-plane $\cal W$ components of the perturbing acceleration $\delta {\bm a}$ respectively by
${\cal R} \equiv {\bm n} \cdot \delta {\bm a}$,
 ${\cal S} \equiv {\bm \lambda} \cdot \delta {\bm a}$ and
 ${\cal W} \equiv \bm{\hat{h}} \cdot \delta {\bm a} = 0$,
and write down the ``Lagrange planetary equations'' for the evolution of the orbit elements,
\begin{eqnarray}
\frac{dp}{dt} &=& 2 \sqrt{\frac{p}{GM}} \,r {\cal S}\,,
\nonumber \\
\frac{d\alpha}{dt} &=& \sqrt{\frac{p}{GM}} \left [ {\cal R}  \sin \phi   +  {\cal S} (\alpha + \cos \phi) \left ( 1+ \frac{r}{p} \right ) 
\right ]\,,
\nonumber \\
\frac{d\beta}{dt} &=& \sqrt{\frac{p}{GM}} \left [ -{\cal R}  \cos \phi   +  {\cal S} (\beta + \sin \phi) \left ( 1+ \frac{r}{p} \right ) 
 \right ]\,,
\label{eq2:Lagrange}
\end{eqnarray}
along with the relation $d\phi/dt = h/r^2$, which can be used to convert $d/dt$ to $d/d\phi$ in the equations
(see \cite{PW2014} for further discussion).  At lowest order (no perturbations), the elements $p$, $\alpha$, and $\beta$ are constant; those solutions can be plugged into the right-hand side, the equations integrated to find corrections, and so on.  The corrections to the orbit elements tend to be of two classes: {\em periodic} corrections, which vary on an orbital time scale, and {\em secular} corrections, which vary on a longer time scale, depending upon the nature of the perturbations.   We adopt the ``multiple-scale'' approach to analyse these corrections systematically to high PN orders.

The Lagrange planetary equations for the orbit elements $X_\alpha$ can be expressed as differential equations in $\phi$, in the general form
\begin{equation}
\frac{d X_\alpha (\phi)}{d\phi} = \epsilon (r^2/h) Q_\alpha (X_\beta(\phi), \phi) \,,
\label{eq2:dXdf}
\end{equation}
where $\alpha, \, \beta$ label the orbit element, $\epsilon$ is a small parameter that characterizes the perturbation, and the $Q_\alpha$ denote the right-hand sides of Eqs.\ (\ref{eq2:Lagrange}). 

In a two-time-scale analysis \cite{1978amms.book.....B,1990PhRvD..42.1123L,2004PhRvD..69j4021M,2008PhRvD..78f4028H}, we assume that the $X_\alpha$ have pieces that vary on a ``short'' orbital time scale, corresponding to the periodic functions of $\phi$, but may also have pieces that vary on a long time scale, of order $1/\epsilon$ times the short time scale.  We treat these two times formally as independent variables, and solve the ordinary differential equations as if they were partial differential equations for the two variables.
We define the long-time-scale variable
$\theta \equiv \epsilon \phi$ and write the derivative with respect to $\phi$ as
$d/d\phi \equiv \epsilon \partial/\partial \theta + \partial/\partial \phi$. 
We make an {\em ansatz} for the solution for $X_\alpha (\theta, \phi)$:
\begin{equation}
X_\alpha (\theta, \phi) \equiv \tilde{X}_\alpha (\theta) + \epsilon Y_\alpha (\tilde{X}_\beta (\theta), \phi) \,,
\label{eq2:ansatz}
\end{equation}
where
\begin{equation}
\tilde{X}_\alpha (\theta) = \langle X_\alpha (\theta, \phi) \rangle \,, \quad \langle Y_\alpha (\tilde{X}_\beta (\theta), \phi) \rangle = 0 \,,
\label{eq2:split}
\end{equation}
where the ``average'' $\langle \dots \rangle$ is defined by 
\begin{equation}
\langle A \rangle \equiv \frac{1}{2\pi} \int_0^{2\pi} A(\theta,\phi) d\phi \,,
\label{eq2:averagedef}
\end{equation}
holding $\theta$ fixed.  
For any function $A(\theta,\phi)$ we define the ``average-free'' part as
${\cal AF}(A) \equiv  A(\theta,\phi) - \langle A \rangle$.
Applying this split to Eq.\ (\ref{eq2:dXdf}) and
taking the average and average-free parts, we obtain
\begin{eqnarray}
\frac{d\tilde{X}_\alpha}{d\theta} &=& \langle (r^2/h) Q_\alpha (\tilde{X}_\beta + \epsilon Y_\beta, \phi) \rangle \,,
\label{eq2:aveq}\\
\frac{\partial Y_\alpha}{\partial \phi} &=& {\cal AF} \left ((r^2/h) Q_\alpha (\tilde{X}_\beta + \epsilon Y_\beta, \phi) \right )  - \epsilon \frac{\partial Y_\alpha}{\partial \tilde{X}_\gamma} \frac{d\tilde{X}_\gamma}{d\theta} \,,
\label{eq2:avfreeeq}
\end{eqnarray}
where we sum over the repeated index $\gamma$.
These equations can then be iterated in a straightforward way.  At zeroth order, Eq.\ (\ref{eq2:aveq}) yields $d\tilde{X}_\alpha/d\theta = \langle Q^0_\alpha  \rangle$ where $Q^0_\alpha \equiv Q_\alpha (\tilde{X}_\beta, \phi)$, which is the conventional result whereby one averages the perturbation holding the orbit elements fixed.  We write the expansion $Y_\alpha  \equiv Y^{(0)}_\alpha + \epsilon Y^{(1)}_\alpha + \epsilon^2 Y^{(2)}_\alpha + \dots$.  We then integrate Eq.\ (\ref{eq2:avfreeeq}) holding $\theta$ fixed to obtain $Y^0_\alpha$.  The iteration continues until we obtain all contributions to  $d\tilde{X}_\alpha/d\theta$ compatible with the order in $\epsilon$ to which $Q_\alpha$ is known, 3PN order in this case.  Note that, when going to higher orders, it is essential to feed the periodic, average-free terms back into the equation for the evolution of the averaged orbit elements.  The final solution including periodic terms is given by Eq.\ (\ref{eq2:ansatz}), with the secular evolution of the $\tilde{X}_\alpha$ given by solutions of Eqs.\ (\ref{eq2:aveq}).  From these solutions one can reconstruct the instantaneous orbit using Eqs.\ (\ref{eq2:keplerorbit}).  

These iterations were carried out in more general contexts  (arbitrary mass ratios, including gravitational radiation, including the spin of the central object) by Mora and Will \cite{2004PhRvD..69j4021M} and Will and Maitra \cite{2017PhRvD..95f4003W}; here we quote only those parts of the results needed for this analysis.
The average-free parts of the perturbed orbit elements were obtained to 1PN, 2PN and 3PN orders; here we quote only the 1PN  terms:
\begin{eqnarray}
Y^0_p &=& -8 \tp \frac{m}{\tp} (\tal \cos \phi + \tbe \sin \phi ) \,,
\nonumber \\
Y^0_\alpha &=& -\frac{m}{2 \tp} \left [  5\tbe \sin 2\phi +5\tal \cos 2\phi  +2 (3+7\tal^2-\tbe^2) \cos \phi  + 16 \tal \tbe \sin \phi \right ] 
\,,
\nonumber \\
Y^0_\beta &=& -\frac{m}{2 \tp} \left [  5\tal \sin 2\phi - 5\tbe \cos 2\phi +2 (3+7\tbe^2-\tal^2) \sin \phi 
+ 16 \tal \tbe \cos \phi \right ] 
\,.
\nonumber \\
\label{eq2:avfree}
\end{eqnarray}

Carrying out the iterations to an order consistent with the 3PN order of the equations of motion, and converting back from $\theta$ to $\phi$, we obtain for the long-timescale evolution of the orbit elements,
\begin{eqnarray}
\frac{d\tp}{d\phi} &=&  0 \,,
\label{eq:dpdtheta}
 \\
\frac{d\tal}{d\phi} &=& -\frac{3m}{\tp} \tbe 
+\frac{3}{4} \left ( \frac{m}{\tp} \right )^2 \tbe \left [ 10 - \tal^2 - \tbe^2  \right ]
\nonumber \\
&& \qquad
- \frac{3}{2}  \left ( \frac{m}{\tp} \right )^{3} \tbe \left [ (29 + 34\tal^2 + 34\tbe^2) \right ]
\,,
\label{eq:daldtheta}
 \\
\frac{d\tbe}{d\phi} &=& \frac{3m}{\tp} \tal 
-\frac{3}{4} \left ( \frac{m}{\tp} \right )^2 \tal \left [ 10 - \tal^2 - \tbe^2  \right ]
\nonumber \\
&& \qquad
+ \frac{3}{2}  \left ( \frac{m}{\tp} \right )^{3} \tal \left [ (29 + 34\tal^2 + 34\tbe^2)  \right ]
\,.
\label{eq:dbedtheta}
\end{eqnarray}
\label{eq:dXdtheta}
From the definitions of $\alpha$ and $\beta$ we obtain $d\te/d\phi = 0$ and
Eq.\ (\ref{eq:domdtheta2}). 

The final solution for the orbital separation as a function of $\phi$ is given by
\begin{equation}
r = \frac{\tp + Y^0_p+ \dots}{1 + (\tal + Y^0_\alpha + \dots) \cos \phi +  (\tbe + Y^0_\beta +\dots) \sin \phi} \,.
\label{eq:r}
\end{equation} 

The 3PN equations of motion admit conservation laws for energy and angular momentum;  quoting selected results to 2PN order from \cite{2017PhRvD..95f4003W}, we have
\begin{eqnarray}
\frac{E}{c^2} &=&1  -\frac{1}{2} \frac{m(1-\te^2)}{ \tp} + \frac{1}{8} \left (\frac{m}{\tp} \right )^2 \left ( 19+38 \te^2 +3\te^4 \right )
\nonumber \\
&&
\quad
 -   \frac{1}{16}\left (\frac{m}{\tp} \right )^3  \left (197 - 205 \te^2 - 98 \te^4 -5 \te^6 \right )
\,,
\nonumber \\
L &=& \sqrt{GM \tp} \biggl \{ 1 + \frac{m}{2\tp} (7 + \te^2) - \frac{1}{8} \left (\frac{m}{\tp} \right )^2 
\left ( 37 -18 \te^2 -3\te^4 \right )
\biggr \}
\,.
\label{eq2:ELconserved}
\end{eqnarray}

As before, we iterate Eqs.\ (\ref{eq2:ELconserved})  to express $\te$ and $\tp$ in terms of $E$ and $L$, and thence in terms of $e_0$ and $p_0$ via Eqs.\ (\ref{eq:epdefinitions}), with the result
\begin{eqnarray}
\te^2 &=& e_0^2 +  x_0 (2 - 15 e_0^2 - 2e_0^4) - \frac{3}{8} x_0^2 ( 56 -74 e_0^2 -57 e_0^4 - 8 e_0^6 ) + O(x_0^3) \,,
\nonumber \\
\tilde{x} &=& x_0 + x_0^2 (7+ e_0^2) + x_0^3 (54 + 7e_0^2) + O(x_0^4) \,,
\label{eq:extransformosc}
\end{eqnarray}
where $\tilde{x} = m/\tp$.  Substituting Eqs.\ (\ref{eq:extransformosc}) into (\ref{eq:domdtheta2}) and expanding to 3PN order yields
\begin{eqnarray}
\frac{d\tom}{d\phi} &= 3 \frac{GM}{c^2 p_0} + \frac{3}{4} \left ( \frac{GM}{c^2 p_0} \right )^{2} \left ( 18 + 5 e_0^2  \right )
+ \frac{3}{2} \left ( \frac{GM}{c^2 p_0} \right )^{3}  \left (136 + 75 e_0^2 \right )  \,.
\label{eq:dtomdthetaosc}
\end{eqnarray}
This appears to be in disagreement with the results of the geodesic method.  This is because the angle between successive pericenters is {\em not} equal to $2\pi$ times the rate of advance $d\tom/d\phi$.  To see this, we solve Eqs.\ (\ref{eq:daldtheta}) and (\ref{eq:dbedtheta}) directly.  Taking into account that $\te$ and $\tp$ are constant over the long timescale and that $\tal^2 + \tbe^2 = \te^2$, and imposing the initial condition that $\tbe/\tal = \tan \omega_0$, we obtain the solutions
\begin{eqnarray}
\tal &=& \te \cos \left ( \tilde{\Omega} \phi + \omega_0 \right ) \,,
\nonumber \\
\tbe &=& \te \sin \left ( \tilde{\Omega} \phi + \omega_0 \right ) \,,
\label{eq:albesolutions}
\end{eqnarray}
where $\tilde{\Omega} = d\tom/d\phi$.  Substituting this into Eq.\ (\ref{eq:r}), we obtain
\begin{equation}
r = \frac{\tp + R}{1 + \te \cos \left [ (1-\tilde{\Omega} ) \phi - \omega_0 \right ] + Q} \,,
\label{eq:r2}
\end{equation} 
where the periodic PN corrections $R$ and $Q$ turn out to be all functions of $\cos N[(1-\tilde{\Omega})\phi - \omega_0]$, where $N \ge 1$ is an integer.  The pericenter or apocenter occur where $dr/d\phi  \propto \sin N[(1-\tilde{\Omega})\phi - \omega_0]  \propto \sin [(1-\tilde{\Omega})\phi - \omega_0] = 0$, which is satisfied by the condition $(1-\tilde{\Omega})\phi - \omega_0 = M\pi$, where $M$ is an integer.  Thus, the change in angle $\Delta \phi$ between successive pericenters, given by $\Delta M=2$, is given by
\begin{equation}
(1 - \tilde{\Omega} ) \Delta \phi = 2\pi \,.
\end{equation}
Thus the advance in the pericenter angle $\Delta \omega$ over one orbit is given
by Eq.\ (\ref{eq:delomega}).  Combining this with Eq.\ (\ref{eq:dtomdthetaosc}) yields Eq.\ (\ref{eq:deltaominv}), in agreement with the other methods.

\section{Discussion}
\label{sec:concluding}

Hoenselaers \cite{1976PThPh..56..324H} analysed the pericenter advance for a test particle in the equatorial plane of the Weyl geometry, which represents a body with multipole moments.  In the spherical limit, which corresponds to the Schwarzschild geometry in unusual coordinates, his result for $\Delta \omega$ in terms of $L$ agrees with Eq.\ (\ref{eq:deltaominv}) to 2PN order.  He did not explicitly relate the eccentricity to $E$, but that only makes a difference at 3PN order. 
Damour and Sch\"afer \cite{1988NCimB.101..127D} and Ohta and Kimura \cite{1989PThPh..81..679O} used 2PN equations of motion to derive $\Delta \omega$ expressed in terms of $E$ and $L$, and  for arbitrary mass ratios.  In the test-body limit, their result agrees with Eq.\ (\ref{eq:deltaominv}) to 2PN order.  Do-Nhat  \cite{1998PhLA..238..328D} used an asymptotic series method to solve the geodesic equation in Schwarzschild coordinates and expressed the pericenter advance in terms of $E$ and $L$, in agreement with Eq.\ (\ref{eq:deltaominv}) to 3PN order.  In his Living Review, Blanchet \cite{2014LRR....17....2B} quotes $\Delta \omega$ to 3PN order in terms of $E$ and $L$ for arbitrary mass ratios.  In the test body limit, his expression matches Eq.\ (\ref{eq:deltaominv}). D'Eliseo \cite{2011Ap&SS.332..121D} and Poveda and Mar\'in 
\cite{2018Ap&SS.363..245M} obtained 3PN corrections using the geodesic method, but they defined eccentricity using none of the methods described here, so their higher PN-order  results differ from each other and from  those displayed in Sec.\ \ref{sec:intro}.   
 
For the most part, our discussion of the pericenter advance in terms of the ``invariant'' quantities $e_0$ and $p_0$ is of theoretical and pedagogical interest, but of little practical impact.  In real-life analyses of orbital data in either the solar system or binary pulsars, a coordinate system is selected, usually the harmonic coordinates of post-Newtonian theory, and all computations are carried out in those coordinates.  The useful observable quantities are not $E$ and $L$, but round trip travel times or Doppler shifts  of radar tracking signals, sky positions measured by VLBI, arrival times of pulsar pulses, and so on, all of which can be related to variables such as the position and velocity of the relevant bodies in harmonic coordinates.  Those variables can then be related to such orbital variables as pericenter angle or eccentricity, all defined in the same coordinates.  As long as this is carried out consistently in the chosen coordinate system, then one can ask such questions as: have we measured the 2PN correction to the rate of pericenter advance as given by Eq.\ (\ref{eq:domdtheta2}) and does it agree with the prediction of general relativity.  Nevertheless, as we begin to explore the dynamics of increasingly relativistic systems, the question of the higher-order contributions to Einstein's famous pericenter advance effect may become relevant.

\ack

This work was supported in part by the US National Science Foundation,
Grant No.\  PHY 16-00188.   AT thanks the Institute for High Energy Physics and Astrophysics and the Institute for Fundamental Theory at the University of Florida for financial support.  We are grateful for the hospitality of  the Institut d'Astrophysique de Paris where much of this work was carried out.    

\section*{References}

\bibliographystyle{iopart-num}

\providecommand{\newblock}{}

\end{document}